\documentclass[11pt]{IEEEtran}
\usepackage{amsmath,amssymb,amsfonts}
\usepackage{algorithmic}
\usepackage{graphicx}
\graphicspath{{./Images/}}
\usepackage{textcomp}
\usepackage{xcolor}
\usepackage{subcaption}
\usepackage{siunitx}
\usepackage{booktabs}
\def\bA{\mathbf{A}}
\def\bB{\mathbf{B}}

\def\bD{\mathbf{D}}
\def\bE{\mathbf{E}}

\def\bG{\mathbf{G}}
\def\bH{\mathbf{H}}

\def\bJ{\mathbf{J}}
\def\bK{\mathbf{K}}

\def\bM{\mathbf{M}}

\def\bS{\mathbf{S}}

\def\bW{\mathbf{W}}

\def\ba{\mathbf{a}}

\def\bc{\mathbf{c}}
\def\bd{\mathbf{d}}

\def\bi{\mathbf{i}}

\def\bm{\mathbf{m}}
\def\bn{\mathbf{n}}

\def\bp{\mathbf{p}}
\def\bq{\mathbf{q}}
\def\br{\mathbf{r}}

\def\bx{\mathbf{x}}

\newcommand{\bnabla}{\mbox{{\boldmath $\nabla$}}}

\newcommand{\btau}{\mbox{{\boldmath $\tau$}}}
\newcommand{\bnu}{\mbox{{\boldmath $\nu$}}}

\usepackage{pgfplots}
\pgfplotsset{compat=newest}
\usepackage{tikz}
\usetikzlibrary{spy}

\begin{document}
\title{Generalized Signal Models and Direct FID-Based Dielectric Parameter Retrieval in MRI}
\author{Patrick S. Fuchs, \IEEEmembership{Student Member, IEEE}, and Rob F. Remis
\thanks{P.S.Fuchs and R.F.Remis are with the Delft University of Technology, Delft, The Netherlands (e-mail: p.s.fuchs@tudelft.nl).}}

\maketitle

\begin{abstract}
In this paper we present full-wave signal models for magnetic and electric field measurements in magnetic resonance imaging (MRI). Our analysis is based on a scattering formalism in which the presence of an object or body is taken into account via an electric scattering source. We show that these signal models can be evaluated, provided the  Green's tensors of the background field are known along with the dielectric parameters of the object and the magnetization within the excited part of the object. Furthermore, explicit signal expressions are derived in case of a small homogeneous ball that is embedded in free space and for which the quasi-static Born approximation can be applied. The conductivity and permittivity of the ball appear as explicit parameters in the resulting signal models and allow us to study the sensitivity of the measured signals with respect to these dielectric parameters. Moreover, for free induction decay signals we show that under certain conditions it is possible to retrieve the dielectric parameters of the ball from noise-contaminated induction decay signals that are based on electric or magnetic field measurements.     
\end{abstract}

\begin{IEEEkeywords}
Magnetic resonance imaging, scattering formalism, Born approximation, free induction decay, dielectric parameter retrieval
\end{IEEEkeywords}

\section{Introduction}

The influence of biological tissue on a typical magnetic resonance imaging (MRI) experiment (and previously in nuclear magnetic resonance (NMR) or zeugmatography~\cite{Hoult&Lauterbur}) has been investigated almost as long as the imaging modality exists. 
Most of this research has focussed on the signal to noise ratio (SNR) of the received signals \cite{hill68}, and on the influence of tissue on the antenna sensitivity patterns \cite{bottomley78}. Both of these aspects play an important role in understanding the structure of the received signal, of course, and are taken into account in signal optimization frameworks as shown in the recent work \cite{lattanzi12}, for example. The influence of scattering currents, induced in biological tissue through the magnetization itself, however is neglected in research on this matter up till now. 

Due to the relationship between the SNR and the MRI background field there is a continuing push to higher field strengths to achieve improved SNRs and faster scan times. These improvements do come at a cost as with higher field strengths also the frequency at which the MRI measurement is performed increases. This higher frequency leads to new challenges in RF coil design, for example, and the received signals are generally more sensitive to changes in the dielectric (tissue) parameters as well. 

In RF coil design a major challenge at higher fields is to achieve a uniform excitation of the region of interest (ROI). Since the size of the object is on the order of the wavelength,  non-uniform RF fields and interference patterns may appear within the ROI. Possible solutions are increasing antenna array sizes and combining antenna types, although it has been demonstrated that such an approach has diminishing returns for larger array sizes~\cite{keil13}. Another approach is varying the array elements, using dipoles~\cite{raaijmakers16-2}, combining loops and dipoles~\cite{ertuerk17, lattanzi18}, or using ``special'' fractionated dipoles~\cite{raaijmakers16}. In most of these approaches the goal is to optimize the so-called ultimate intrinsic signal to noise ratio (UISNR) or, in other words, to approximate ideal current patterns which would lead to the highest SNR~\cite{lattanzi12}. Originally the term UISNR was introduced in \cite{ocali98}, but additions have been made ever since, covering parallel MRI \cite{ohliger03}, current patterns required to attain this ratio \cite{lattanzi12}, and addition of the specific absorption rate (SAR) \cite{kopanoglu11}.

For the SAR all of the above-mentioned challenges are combined, as the higher heterogeneity of the RF fields leads to a local increase in tissue heating, which limits the amount of current that can be used to power a measurement and thus limits the SNR that can be obtained for a specific field strength and antenna array. Validated simulation techniques may be used to obtain more accurate local SAR estimates and may lead to antenna designs with reduced restrictions on the antenna currents that can be employed, or dielectric pads (passive shimming) can be used to improve the field homogeneity and reduce local heating effects \cite{wyger14}. 

In this paper, we focus on the signal modeling part and derive full-wave signal models based on Maxwell's equations. Electric and magnetic field measurements are considered and we show that the resulting signals are due to the time-varying magnetization inside the object and the induced electric scattering currents, each weighted by their own receive field as determined by the coil or antenna that is used for reception. The signal models can be explicitly evaluated provided the Green's tensors of the background medium and the medium parameters of the object are known. Moreover, to gain further insight into how the electromagnetic medium parameters of the object influence the measured signal, explicit time- and frequency-domain signal models are derived for a special case, where the background medium consists of air and the object is a homogeneous ball that is uniformly excited and for which the Born approximation applies. Quasi-static signal representations are derived from the full-wave signal models and through a series of numerical experiments we verify our models for the received signals. Finally, we demonstrate that under certain conditions it is possible to retrieve the dielectric parameters of the ball from measured free induction decay signals that are based on electric or magnetic field measurements.  
   
We present our analysis in the Laplace- or $s$-domain, since it allows us to easily obtain frequency-domain solutions by letting $s \rightarrow \text{j}\omega$ via the right-half of the complex $s$-plane, or time-domain field responses using standard Laplace transformation rules. 
   
\section{Theory} 
Let $\mathbb{D}_{\text{obj}}$ be a bounded domain occupied by a penetrable object that is present in an MR scanner. We assume that the complete object or part of this object has been excited during the transmit state of the scanner. More precisely, we assume that the temporal derivative of magnetization $\partial_t \bM(\bx,t)$ is nonzero within the subdomain $\mathbb{D}_{\text{ex}} \subseteq \mathbb{D}_{\text{obj}}$ and vanishes outside this domain. In other words, $\partial_t \bM(\bx,t)$ has the domain $\mathbb{D}_{\text{ex}}$ as its spatial support and $\mathbb{D}_{\text{ex}}=\mathbb{D}_{\text{obj}}$ if the complete object is excited.

Measurements are carried out outside the object and take place in free space. To set up the data models that describe our measurements, we first consider a surface~$S$ with unit normal $\bnu$ and closed boundary curve $C$ with a unit normal $\btau$ along this curve such that $\btau$ and $\bnu$ are oriented according to the right-hand rule. The surface~$S$ has an area $A$ and the position vector of its barycenter is denoted by $\bx_{\text{R}}$. The surface is completely located in air and is used to measure the electromotive or magnetomotive force given by 
\begin{equation}   
\label{eq:emf_mmf}
\hat{V}_{\text{emf}}(s) = 
\int_{\bx \in C}
\hat{\bE} \cdot \btau \, \text{d}\ell
\quad \text{and} \quad
\hat{I}_{\text{mmf}}(s) = 
\int_{\bx \in C}
\hat{\bH} \cdot \btau \, \text{d}\ell,
\end{equation}
respectively. Using Maxwell's equations and assuming that the area~$A$ of the surface is sufficiently small (diameter much smaller than the smallest wavelength of interest), we have 
\begin{equation}
\label{eq:emf2}
\hat{V}_{\text{emf}}(s) = -s 
\int_{\bx \in S} \hat{\bB} \cdot \bnu \, \text{d}A 
\approx 
-s \mu_0 A \, \hat{\bH}(\bx_{\text{R}},s) \cdot \bnu, 
\end{equation}  
where we have used $\hat{\bB}=\mu_0\hat{\bH}$, since the measurement surface~$S$ is located in air. Similarly, for the magnetomotive force we obtain 
\begin{equation}
\label{eq:mmf2}
\hat{I}_{\text{mmf}}(s) = s 
\int_{\bx \in S} \hat{\bD} \cdot \bnu \, \text{d}A 
\approx 
s \varepsilon_0 A \, \hat{\bE}(\bx_{\text{R}},s) \cdot \bnu,
\end{equation}
where we have used $\hat{\bD}=\varepsilon_0 \hat{\bE}$. Assuming that a measurement is linear and time-invariant, we can generalize our field measurement description to  
\begin{align}
\label{eq:signal_h_s}
\hat{d}_{\text{h}}(s) &=
\int_{\bx \in \mathbb{D}_{\text{rec}}}
\hat{\bm}_{\text{h}}(\bx,s) \cdot \hat{\bH}(\bx,s) \, \text{d}V 
\intertext{and}
\label{eq:signal_e_s}
\hat{d}_{\text{e}}(s) &=
\int_{\bx \in \mathbb{D}_{\text{rec}}}
\hat{\bm}_{\text{e}}(\bx,s) \cdot \hat{\bE}(\bx,s) \, \text{d}V, 
\end{align}
in which a volumetric receiver is used to obtain the measured signals. The receiver is completely located outside the object, occupies the receiver domain $\mathbb{D}_{\text{rec}}$, and its action on the electromagnetic field inside the receiver domain is described by the vectorial receiver functions $\hat{\bm}_{\text{h}}$ and $\hat{\bm}_{\text{e}}$ for magnetic and electric field measurements, respectively. Note that the electro- and magnetomotive forces are special cases of (\ref{eq:signal_h_s}) and (\ref{eq:signal_e_s}). In particular, with 
\begin{equation}
\label{eq:mh_emf}
\hat{\bm}_{\text{h}}(s) = s\mu_0 A \delta(\bx-\bx_{\text{R}}) \bnu,
\end{equation}
and 
\begin{equation}
\label{eq:me_mmf}
\hat{\bm}_{\text{e}}(s) = s\varepsilon_0 A \delta(\bx-\bx_{\text{R}}) \bnu,
\end{equation}
we have $\hat{d}_{\text{h}}(s)=-\hat{V}_{\text{emf}}(s)$ and $\hat{d}_{\text{e}}(s)=\hat{I}_{\text{mmf}}(s)$. Since an electromotive force measurement is characterized by (\ref{eq:signal_h_s}) and (\ref{eq:mh_emf}), we refer to such a measurement as a magnetic field measurement, while a magnetomotive force measurement is refered to as an electric field measurement, since it can be described by (\ref{eq:signal_e_s}) and (\ref{eq:me_mmf}). Below, we take the general signal models (\ref{eq:signal_h_s}) and (\ref{eq:signal_e_s}) as a starting point and consider the electro- and magnetomotive forces as special cases. 

\subsection{Scattering Formalism}
To further develop the signal models (\ref{eq:signal_h_s}) and (\ref{eq:signal_e_s}), the magnetic and electric field strengths inside the receiver domain are obviously required. To this end, we set up a scattering formalism and write the electromagnetic field as a superposition of a background and a scattered field. The background field is defined as the field that is present when the constitutive parameters within the object domain are the same as the parameters of the background medium, while the scattered field takes the presence of the object into account. Assuming that the background can be accurately described by a background conductivity $\sigma_{\text{b}}(\bx)$, a background permittivity $\varepsilon_{\text{b}}(\bx)$, and a permeability $\mu_{\text{b}}(\bx)$, the Laplace-domain background field satisfies the Maxwell equations 
\begin{align}
\label{eq:Ms_b1}
-\bnabla \times \hat{\bH}^{\text{b}} + \sigma_{\text{b}} \hat{\bE}^{\text{b}} 
+ s \varepsilon_{\text{b}} \hat{\bE}^{\text{b}} &= \mathbf{0},
\intertext{and}
\label{eq:Ms_b2}
\bnabla \times \hat{\bE}^{\text{b}} + s\mu_{\text{b}} \hat{\bH}^{\text{b}} &= - \hat{\bK},
\end{align}
where $\hat{\bK}$ is the Laplace transform of $\mu_0 \partial_t \bM$ with $\bM(\bx,t)$ the time-varying magnetization with the domain $\mathbb{D}_{\text{ex}}$ as its spatial support. Across interfaces where the background medium parameters exhibit a jump, the above Maxwell's equations have to be supplemented by the appropriate boundary conditions and if perfectly conducting structures are present in the background configuration, then the boundary condition for a perfectly conducting structure has to be included as well, of course. For general inhomogeneous background configurations that can be described in terms of the background medium parameters, the above Maxwell equations can only be solved numerically. Formally, however, we can express the electromagnetic background field in terms of the Green's tensors of the background medium as 
\begin{align}
\label{eq:Hb}
\hat{\bH}^{\text{b}}(\bx,s) = 
\int_{\bx' \in \mathbb{D}_{\text{ex}}} \underline{\hat{\bG}}^{\text{HK}}(\bx,\bx',s) \cdot \hat{\bK}(\bx',s) \, \text{d}V
\intertext{and} 
\label{eq:Eb}
\hat{\bE}^{\text{b}}(\bx,s) = 
\int_{\bx' \in \mathbb{D}_{\text{ex}}} \underline{\hat{\bG}}^{\text{EK}}(\bx,\bx',s) \cdot \hat{\bK}(\bx',s) \, \text{d}V, 
\end{align}
where $\underline{\hat{\bG}}^{\text{HK}}$ and $\underline{\hat{\bG}}^{\text{EK}}$ are the magnetic current to magnetic field and magnetic current to electric field Green's tensors of the background medium. 

Furthermore, the scattered field $\{\hat{\bH}^{\text{sc}}, \hat{\bE}^{\text{sc}}\}$ satisfies the Maxwell equations
\begin{align}
\label{eq:Ms_sc1}
-\bnabla \times \hat{\bH}^{\text{sc}} + \sigma_{\text{b}} \hat{\bE}^{\text{sc}} 
+ s \varepsilon_{\text{b}} \hat{\bE}^{\text{sc}} = -\hat{\bJ}^{\text{sc}}
\intertext{and}
\label{eq:Ms_sc2}
\bnabla \times \hat{\bE}^{\text{sc}} + s\mu_{\text{b}} \hat{\bH}^{\text{sc}} = \mathbf{0},
\end{align}
where $\hat{\bJ}^{\text{sc}}$ is the Laplace transformed dielectric scattering source given by
\begin{equation}
\label{eq:Jsc_s}
\hat{\bJ}^{\text{sc}}(\bx,s) = 
\big\{\sigma(\bx)-\sigma_\text{b}(\bx) +s[\varepsilon(\bx) - \varepsilon_\text{b}(\bx)] \big\} \hat{\bE}(\bx,s) 
\end{equation}
for $\bx \in \mathbb{D}_{\text{obj}}$, where $\sigma(\bx)$ is the conductivity of the object and $\varepsilon(\bx)$ its permittivity. The object is assumed to have no contrast in its permeability with respect to the background medium. 
  
For the scattered field we have the integral representations 
\begin{align}
\label{eq:Hsc}
\hat{\bH}^{\text{sc}}(\bx,s) &= 
\int_{\bx' \in \mathbb{D}_{\text{obj}}} \underline{\hat{\bG}}^{\text{HJ}}(\bx,\bx',s) \cdot \hat{\bJ}^{\text{sc}}(\bx',s) \, \text{d}V \\
\intertext{and}
\label{eq:Esc}
\hat{\bE}^{\text{sc}}(\bx,s) &= 
\int_{\bx' \in \mathbb{D}_{\text{obj}}} \underline{\hat{\bG}}^{\text{EJ}}(\bx,\bx',s) \cdot \hat{\bJ}^{\text{sc}}(\bx',s) \, \text{d}V, 
\end{align}
where $\underline{\hat{\bG}}^{\text{HJ}}$ and $\underline{\hat{\bG}}^{\text{EJ}}$ are the electric current to magnetic field and electric current to electric field Green's tensors of the background medium.  Having the integral representations for the background and scattered fields at our disposal, we can now further develop the full-wave signal models (\ref{eq:signal_h_s}) and (\ref{eq:signal_e_s}). 

\subsection{Full-Wave Signal Model}
Writing the total magnetic and electric fields in the receiver domain as a superposition of the background and scattered fields and using the integral representations (\ref{eq:Hb}), (\ref{eq:Eb}), (\ref{eq:Hsc}), and (\ref{eq:Esc}), the signal models of  (\ref{eq:signal_h_s}) and (\ref{eq:signal_e_s}) become
\begin{align}
\label{eq:sm_dh1}
\begin{split}
&\hat{d}_{\text{h}}(s) = \\
&\int\limits_{\bx \in \mathbb{D}_{\text{ant}}}
\hat{\bm}_{\text{h}}(\bx,s) \cdot \int\limits_{\bx' \in \mathbb{D}_{\text{ex}}} \underline{\hat{\bG}}^{\text{HK}}(\bx,\bx',s) \cdot \hat{\bK}(\bx',s) \, \text{d}V \, \text{d}V \\
&+ \hspace{-0.35cm}
\int\limits_{\bx \in \mathbb{D}_{\text{ant}}}
\hat{\bm}_{\text{h}}(\bx,s) \cdot \int\limits_{\bx' \in \mathbb{D}_{\text{obj}}} \underline{\hat{\bG}}^{\text{HJ}}(\bx,\bx',s) \cdot \hat{\bJ}^{\text{sc}}(\bx',s) \, \text{d}V \, \text{d}V 
\end{split}
\end{align}
and 
\begin{align}
\label{eq:sm_de1}
\begin{split}
&\hat{d}_{\text{e}}(s) = \\
&\int\limits_{\bx \in \mathbb{D}_{\text{ant}}}
\hat{\bm}_{\text{e}}(\bx,s) \cdot  \int\limits_{\bx' \in \mathbb{D}_{\text{ex}}} \underline{\hat{\bG}}^{\text{EK}}(\bx,\bx',s) \cdot \hat{\bK}(\bx',s) \, \text{d}V \, \text{d}V \\
&+ \hspace{-0.35cm}
\int\limits_{\bx \in \mathbb{D}_{\text{ant}}}
\hat{\bm}_{\text{e}}(\bx,s) \cdot \int\limits_{\bx' \in \mathbb{D}_{\text{obj}}} \underline{\hat{\bG}}^{\text{EJ}}(\bx,\bx',s) \cdot \hat{\bJ}^{\text{sc}}(\bx',s) \, \text{d}V \, \text{d}V. 
\end{split}
\end{align}
Interchanging the order of integration and using the reciprocity properties of the Green's tensors \cite{DeHoop} allows us to write the signal representations as 
\begin{align}
\label{eq:dh_final_s}
\begin{split}
\hat{d}_{\text{h}}(s) &=
\int\limits_{\bx' \in \mathbb{D}_{\text{ex}}}
\hat{\bK}(\bx',s) \cdot \hat{\bW}_{\text{h}}^{\text{mg}}(\bx',s) \, \text{d}V \\ 
&-
\int\limits_{\bx' \in \mathbb{D}_{\text{obj}}}
\hat{\bJ}^{\text{sc}}(\bx',s) \cdot \hat{\bW}_{\text{e}}^{\text{mg}}(\bx',s) \, \text{d}V 
\end{split}
\end{align}
and
\begin{align}
\label{eq:de_final_s}
\begin{split}
\hat{d}_{\text{e}}(s) =
&-\int\limits_{\bx' \in \mathbb{D}_{\text{ex}}}
\hat{\bK}(\bx',s) \cdot \hat{\bW}_{\text{h}}^{\text{el}}(\bx',s) \, \text{d}V \\ 
&+ \int\limits_{\bx' \in \mathbb{D}_{\text{obj}}}
\hat{\bJ}^{\text{sc}}(\bx',s) \cdot \hat{\bW}_{\text{e}}^{\text{el}}(\bx',s) \, \text{d}V, 
\end{split}
\end{align}
where we have introduced the receive fields for a magnetic field measurement as  
\begin{align}
	\label{eq:defWh_mg}
 	\hat{\bW}_\text{h}^\text{mg}(\bx',s) &= \int\limits_{\bx \in \mathbb{D}_{\text{rec}}}
 \underline{\hat{\bG}}^{\text{HK}}(\bx',\bx,s) \cdot \hat{\bm}_{\text{h}}(\bx,s)\, \text{d}V,
 \intertext{and}
 \label{eq:defWe_mg}
 		\hat{\bW}_\text{e}^\text{mg}(\bx',s) &= \int\limits_{\bx \in \mathbb{D}_{\text{rec}}}
 \underline{\hat{\bG}}^{\text{EK}}(\bx',\bx,s) \cdot \hat{\bm}_{\text{h}}(\bx,s)\, \text{d}V,
 \end{align}
while the receive fields for an electric field measurement are given by 
 \begin{align}
	\label{eq:defWh_el}
 	\hat{\bW}_\text{h}^\text{el}(\bx',s) &= \int\limits_{\bx \in \mathbb{D}_{\text{rec}}}
 \underline{\hat{\bG}}^{\text{HJ}}(\bx',\bx,s) \cdot \hat{\bm}_{\text{e}}(\bx,s)\, \text{d}V,
 \intertext{and}
 \label{eq:defWe_el}
 		\hat{\bW}_\text{e}^\text{el}(\bx',s) &= \int\limits_{\bx \in \mathbb{D}_{\text{rec}}}
 \underline{\hat{\bG}}^{\text{EJ}}(\bx',\bx,s) \cdot \hat{\bm}_{\text{e}}(\bx,s)\, \text{d}V,
 \end{align}
Equations (\ref{eq:dh_final_s}) and (\ref{eq:de_final_s}) are the full-wave signal models for a magnetic and electric field measurement, respectively, in which the magnetic-current source (magnetization) and the scattering source contribute to the measured signal both weighted by their respective antenna receive fields. To evaluate these models, first the magnetization (and hence the magnetic-current source $\hat{\bK}$) must be known within the excited part $\mathbb{D}_{\text{ex}}$ of the object, since the time variations of this field quantity generate the radiated electromagnetic field. Second, the conductivity and permittivity profiles of the object must be known. This allows us to determine the electric field strength within the object by solving a forward problem with the magnetic-current density $\hat{\bK}$ in $\mathbb{D}_{\text{ex}}$ as a source. Finally, the Green's tensors   
of the background medium must be known as well to determine the receive fields (\ref{eq:defWh_mg}) -- (\ref{eq:defWe_el}). In general, these tensors can only be determined through simulations, since the background is inhomogeneous. In conclusion, the full-wave signals can be evaluated in principle, provided that (i)~the magnetization in $\mathbb{D}_{\text{ex}}$ is known, (ii)~the conductivity and permittivity profiles of the object are known, and (iii)~the Green's tensors of the background medium are known. Note that frequency-domain responses are obtained by letting $s \rightarrow \text{j}\omega$ and time-domain signal responses involve temporal convolutions of the magnetic-current source and the dielectric scattering source with their respective receive fields, since their Laplace-domain counterparts all are $s$-dependent in general.

\subsection{Simplified Full-Wave Signal Models for a Ball Located in Free-Space}
Given the above observations, we consider a specific configuration for which it is possible to develop signal models that explicitly show how the received signals depend on the conductivity and permittivity of the object. In particular, we first consider a background medium consisting of free space. The Green's tensors of the background medium and the receive fields for electro- or magnetomotive force measurements (dipole measurements) can then be determined explicitly. Second, we take a small homogeneous ball with a constant conductivity $\sigma$ and permittivity $\varepsilon$ as our object of interest. Explicit signal models can then be developed provided the radius of the ball is sufficiently small. 
 
Let the background medium be free space and consider an electro- or magnetomotive force measurement. For an electromotive force measurement, the receive function $\hat{\bm}_{\text{h}}$ is given by (\ref{eq:mh_emf}) and since the background medium is free space, the Green's tensors are explicitly known \cite{DeHoop} and the receive fields follow as 
 \begin{align}
\label{eq:Wh_emf_vac}
\begin{split}
&\hat{\bW}_{\text{h}}^{\text{mg}}(\bx',s) = s\mu_0 A \underline{\hat{\bG}}^{\text{HK}}(\bx',\bx_{\text{R}},s) \cdot \bnu \\
&= \frac{A}{4\pi |\bx'-\bx_{\text{R}}|^3} \exp(-s\tau) 
\left[
(1+s\tau) \bp_1 + (s\tau)^2 \bp_{2}
\right]
\end{split}
\end{align}
and
\begin{align}
\label{eq:We_emf_vac}
\begin{split}
&\hat{\bW}_{\text{e}}^{\text{mg}}(\bx',s) = s\mu_0 A \underline{\hat{\bG}}^{\text{EK}}(\bx',\bx_{\text{R}},s) \cdot \bnu \\
&= \frac{s\mu_0 A}{4\pi |\bx'-\bx_{\text{R}}|^2} \exp(-s\tau) 
(1+ s\tau) \bn  \times \bnu,
\end{split}
\end{align}
where $\tau=c_0^{-1}|\bx'-\bx_{\text{R}}|$ with $c_0$ is the electromagnetic wave speed in vacuum. Clearly, $\tau$ is the travel time from the point of integration $\bx'$ to the receiver location $\bx_{\text{R}}$. Furthermore, $\bp_1=3\bn (\bn \cdot \bnu) - \bnu$, and $\bp_2 =\bn (\bn \cdot \bnu) - \bnu$ with $\bn=(\bx'-\bx_{\text{R}})/|\bx'-\bx_{\text{R}}|$ the unit vector pointing from the receiver location to the point of integration.  

Similarly, for a magnetomotive force measurement, the receive function $\hat{\bm}_{\text{e}}$ is given by (\ref{eq:me_mmf}) and the receive fields follow as 
\begin{align}
\label{eq:Wh_mmf_vac}
\begin{split}
&\hat{\bW}_{\text{h}}^{\text{el}}(\bx,s) = s\varepsilon_0 A \underline{\hat{\bG}}^{\text{HJ}}(\bx,\bx_{\text{R}},s) \cdot \bnu \\
&= -\frac{s\varepsilon_0 A}{4\pi |\bx'-\bx_{\text{R}}|^2} \exp(-s\tau) 
(1+ s\tau) \bn  \times \bnu,
\end{split}
\end{align}
and
 \begin{align}
\label{eq:We_mmf_vac}
\begin{split}
&\hat{\bW}_{\text{e}}^{\text{el}}(\bx,s) = s\varepsilon_0 A \underline{\hat{\bG}}^{\text{EJ}}(\bx,\bx_{\text{R}},s) \cdot \bnu \\
&= \frac{A}{4\pi |\bx'-\bx_{\text{R}}|^3} \exp(-s\tau) 
\left[
(1+s\tau) \bp_1 + (s\tau)^2 \bp_{2}
\right].
\end{split}
\end{align}
Note that $\hat{\bW}_{\text{e}}^{\text{mg}}$ and $\hat{\bW}_{\text{h}}^{\text{el}}$ are proportional to each other and $\hat{\bW}_{\text{h}}^{\text{mg}} = \hat{\bW}_{\text{e}}^{\text{el}} $.

Second, we take a small ball as our object of interest. The ball is centered at the origin of our reference frame and has a radius $a>0$. It is characterized by a constant conductivity $\sigma$ and permittivity $\varepsilon$, and its permeability is equal to that of free space. We assume that the radius $a$ is so small that the ball is excited throughout ($\mathbb{D}_{\text{ex}}=\mathbb{D}_{\text{obj}}$) and time variations of the magnetization (and hence the magnetic-current source $\hat{\bK}$) are uniform, that is, $\hat{\bK}$ does not vary with position within the ball. For a given magnetization, the magnetic-current source is now known and the total electric field within the ball can be computed by solving the integral equation
\begin{align}
\label{eq:intE}
\begin{split}
\hat{\bE}(\bx',s) &= \hat{\bE}^{\text{b}}(\bx',s) \\
&- \hat{\chi}  (\hat{\gamma}^2_{0} -\bnabla \, \bnabla \cdot )
\int_{\bx' \in \mathbb{D}_{\text{obj}}} 
\hspace{-0.5cm} \hat{G}(\bx-\bx',s) \hat{\bE}(\bx',s) \, \text{d}V, 
\end{split}
\end{align}
for the electric field $\hat{\bE}(\bx',s)$ with $\bx' \in \mathbb{D}_{\text{obj}}$. In the above equation, $\hat{\gamma}_0=s/c_0$ is the propagation coefficient of free space, $\hat{\chi} = \varepsilon_{\text{r}}-1 + \sigma/(s\varepsilon_0)$ is the contrast of the ball, $\hat{G}$ is the scalar Green's function of free space, and $\hat{\bE}^{\text{b}}$ can be determined from (\ref{eq:Eb}), since $\hat{\bK}$ is known. In the next section, we will essentially follow such an approach, except that we will determine the electric field in the time-domain using FDTD for a given magnetization. Here, we use the above integral equation to arrive at the desired signal models. Specifically, let us consider frequencies of operation $s$ and a ball of radius $a$ with conductivity and permittivity values $\sigma$ and $\varepsilon$, respectively, such that the condition 
\begin{equation}
\label{eq:cond}
(2a |\hat{\gamma_0}|)^2 | \hat{\chi} | \ll1
\end{equation}
is satisfied.  For three-dimensional scalar wave field problems, this is a sufficient condition for the Neumann series to converge~\cite{DeHoop2, Chew}. In addition, let us assume that there is (essentially) no charge accumulation at the boundary of the ball. The gradient-divergence term is then negligible and the above integral equation turns into a scalar integral equation for the electric field. Moreover, since we consider frequencies and dielectric parameters for which (\ref{eq:cond}) holds, we may approximate 
\begin{equation}
\label{eq:Born} 
\hat{\bE}(\bx',s) \approx \hat{\bE}^{\text{b}}(\bx',s).
\end{equation}
Now provided the quasi-static condition $|2a\hat{\gamma}_0 | \ll1$ is also satisfied, this background field is essentially given by
\begin{equation}
\label{eq:Eb_QS}
\hat{\bE}^{\text{b}}(\bx',s) = -\frac{1}{3} \hat{\bK} \times \bx'  
\end{equation}
with $\bx'\in \mathbb{D}_{\text{obj}}$. Notice that this background field does not have a radial component, which is consistent with our assumption of no charge accumulation at the boundary. Also note that if the quasi-static condition $|2a\hat{\gamma}_0 | \ll1$ holds, then (\ref{eq:cond}) can be satisfied for  $|\hat{\chi} | \gg 1$ \cite{Chew}.  

Provided the quasi-static and Born approximation hold, the dielectric scattering source within the ball is given by  
\begin{align}
\label{eq:sc_src_Born}
\begin{split}
\hat{\bJ}^{\text{sc}}(\bx',s) &= 
[\sigma + s(\varepsilon-\varepsilon_0)] \hat{\bE}(\bx',s) \\
&= 
-\frac{1}{3}
[\sigma + s(\varepsilon-\varepsilon_0)] \hat{\bK}(s) \times \bx', 
\end{split}
\end{align}
for $\bx'\in \mathbb{D}_{\text{obj}}$. Substitution in (\ref{eq:dh_final_s}) and (\ref{eq:de_final_s}), we obtain the signal models
\begin{equation}
\label{eq:dh_Born}
\hat{d}_{\text{h}}(s) = 
\hat{\bK}(s) \cdot
\int_{\bx'\in \mathbb{D}_{\text{obj}}}
\hat{\bS}^{\text{mg}}(\bx',s) \, \text{d}V
\end{equation}
and 
\begin{equation}
\label{eq:de_Born}
\hat{d}_{\text{e}}(s) = 
-\hat{\bK}(s) \cdot
\int_{\bx'\in \mathbb{D}_{\text{obj}}}
\hat{\bS}^{\text{el}}(\bx',s) \, \text{d}V,
\end{equation}
where the vectorial sensitivity functions are given by 
\begin{equation}
\label{eq:defS_mg}
\hat{\bS}^{\text{mg}}(\bx',s) = 
\hat{\bW}^{\text{mg}}_{\text{h}}+ 
\hat{\chi}_{\text{e}}
\bx'\times \hat{\bW}^{\text{mg}}_{\text{e}}
\end{equation}
and
\begin{equation}
\label{eq:defS_el}
\hat{\bS}^{\text{el}}(\bx',s) = 
\hat{\bW}^{\text{el}}_{\text{h}}+ 
\hat{\chi}_{\text{e}}
\bx'\times \hat{\bW}^{\text{el}}_{\text{e}}
\end{equation}
with $\hat{\chi}_{\text{e}} = [\sigma + s(\varepsilon-\varepsilon_0)]/3=s\varepsilon_0\hat{\chi}/3$. Substituting expressions (\ref{eq:Wh_emf_vac}) -- (\ref{eq:We_mmf_vac}) for the receive fields in the above equations and applying an inverse Laplace transform ,we obtain the time-domain signals 
\begin{equation}
\label{eq:dh_Born_td}
d_{\text{h}}(t) = 
\mu_0 \partial_t \bM(t) \overset{t}{\ast}  \int_{\bx' \in \mathbb{D}_{\text{obj}}} \bS^{\text{mg}}(\bx',t-\tau)\,\mathrm{d}V,
\end{equation}
and 
\begin{equation}
\label{eq:de_Born_td}
d_{\text{e}}(t) = 
-\mu_0 \partial_t \bM(t) \overset{t}{\ast}  \int_{\bx' \in \mathbb{D}_{\text{obj}}} \bS^{\text{el}}(\bx',t-\tau)\,\mathrm{d}V,
\end{equation}
for $t>0$, where the asterisk denotes convolution in time and the time-domain sensitivity functions are given by
\begin{equation}
\label{eq:Smg_td}
\bS^{\text{mg}}(\bx',t) = 
\frac{A}{4\pi|\bx'- \bx_{\text{R}}|^3}
\sum_{k=0}^{3} 
\tau^{k} \delta^{(k)}(t) \br_{k}^{\text{mg}}
\end{equation}
and 
\begin{equation}
\label{eq:Sel_td}
\bS^{\text{el}}(\bx',t) = 
\frac{A}{4\pi|\bx'- \bx_{\text{R}}|^3}
\sum_{k=0}^{3} 
\tau^{k} \delta^{(k)}(t) \br_{k}^{\text{el}},
\end{equation}
where $\delta^{(k)}$ is the $k$th derivative of the Dirac distribution. Explicit expressions for the expansion vectors $\br_{k}^{\text{mg,el}}$, $k=0,1,2,3$, are given in the Appendix.

In the above signal models, propagation effects and travel times from the ball to the receiver are fully taken into account. However, when the receivers are located not too far from the ball (in a sense to be made precise) then the signals may be simplified even further.  To this end, we substitute the receive fields of (\ref{eq:Wh_emf_vac}) -- (\ref{eq:We_mmf_vac}) in (\ref{eq:defS_mg}) and (\ref{eq:defS_el}) and arrange the resulting expressions in such a way that the sensitivities are expanded in terms of vectors that do not depend on the distance $|\bx'- \bx_{\text{R}}|$. Carrying out these steps, we find for the magnetic field sensitivity function 
\begin{align}
\label{eq:S_mg_fd}
\begin{split}
&\hat{\bS}^{\text{mg}}(\bx',s) = 
\frac{A}{4\pi |\bx'- \bx_{\text{R}}|^3} \exp(-s\tau) \cdot\\ 
&\left[
\bp_1 + 
(s\tau)(\bp_1 + \hat{\bq}^{\text{mg}}) 
+(s\tau)^2 (\bp_2 + \hat{\bq}^{\text{mg}})\right],
\end{split}
\end{align}
while for the sensitivity function for an electric field measurement, we have 
\begin{align}
\label{eq:S_el_fd}
\begin{split}
\hat{\bS}^{\text{el}}(\bx',s) &=
\frac{A}{4\pi |\bx'- \bx_{\text{R}}|^3} \exp(-s\tau) \cdot\\ 
&\big[
\hat{\chi}_{\text{e}} \bx'\times \bp_1 \\
&+ (s\tau)(\hat{\chi}_{\text{e}} \bx'\times \bp_1 + Y_0\bq^{\text{el}})  \\
&+ (s\tau)^2 (\hat{\chi}_{\text{e}} \bx' \times \bp_2 + Y_0\bq^{\text{el}})\big]
\end{split}
\end{align}
with $\hat{\bq}^{\text{mg}} =  Z_0\hat{\chi}_{\text{e}} [(\bx' \cdot \bnu) \bn - (\bx'\cdot \bn)\bnu]$ and $\bq^{\text{el}} =  \bnu \times \bn$.
Note that the vectors 
\begin{equation}
\label{eq:expv}
\bp_{1,2} + \hat{\bq}^{\text{mg}}
\quad \text{and} \quad  
\hat{\chi}_{\text{e}} \bx'\times \bp_{1,2} + Y_0\bq^{\text{el}}
\end{equation}
are $s$-dependent, but do not depend on $|\bx'-\bx_{\text{R}}|$. We can now use the above expressions to investigate which terms contribute to the received signals measured at different receiver locations. Specifically, let us first consider the case where we place the receiver near (almost at) the surface of the ball ($|\bx_{\text{R}}|=a(1+  \epsilon)$, with $\epsilon>0$ small). In this case, $|s|\tau \leq |\hat{\gamma}_0 2a |\ll 1$ and the receive field can be considered quasi-static. The signals models simplify to 
    \begin{align}
\label{eq:dh_Born_nearQS}
\begin{split}
\hat{d}_{\text{h}}(s) &\approx \hat{d}_{\text{h}}^{\text{QS}}(s) \\
&=
\frac{A}{4\pi} 
\left[ \hat{\bK} \cdot 
\int_{\bx'\in \mathbb{D}_{\text{obj}}}
\frac{\bp_1}{|\bx'-\bx_{\text{R}}|^3} \, \text{d}V \right. \\
+&\left. s\mu_0 \hat{\chi}_{\text{e}} \hat{\bK} \cdot
\int_{\bx'\in \mathbb{D}_{\text{obj}}}
\frac{(\bx'\cdot \bnu)\bn - (\bx'\cdot \bn)\bnu}{|\bx'-\bx_{\text{R}}|^2} \, \text{d}V \right]
\end{split}
\end{align}
and 
\begin{align}
\label{eq:de_Born_nearQS}
\begin{split}
\hat{d}_{\text{e}}(s) &\approx \hat{d}_{\text{e}}^{\text{QS}}(s) \\
&=
-\frac{A}{4\pi} 
\left[\hat{\chi}_{\text{e}} \hat{\bK} \cdot 
\int_{\bx'\in \mathbb{D}_{\text{obj}}}
\frac{\bx'\times \bp_1}{|\bx'-\bx_{\text{R}}|^3} \, \text{d}V \right. \\
+&\left. s\varepsilon_0  \hat{\bK} \cdot
\int_{\bx'\in \mathbb{D}_{\text{obj}}}
\frac{\bnu \times \bn}{|\bx'-\bx_{\text{R}}|^2} \, \text{d}V \right]
\end{split}
\end{align}
and their time-domain counterparts are given by 
\begin{align}
\label{eq:dh_Born_td_QS}
\begin{split}
&d_{\text{h}}^{\text{QS}}(t) = 
\frac{\mu_0 A}{4\pi}
\left[  \partial_t \bM \cdot 
\int_{\bx' \in \mathbb{D}_{\text{obj}}} 
\frac{\bp_1}{|\bx'-\bx_{\text{R}}|^3}\,\mathrm{d}V \right .\\
&+ 
\frac{\sigma\mu_0}{3} \partial_t^2 \bM \cdot 
\int_{\bx' \in \mathbb{D}_{\text{obj}}} 
\frac{[(\bx'\cdot \bnu)\bn - (\bx'\cdot \bn)\bnu]}{|\bx'- \bx_{\text{R}}|^2} \, \text{d}V  \\
&+ \left. 
\frac{\varepsilon_{\text{r}}-1}{3} \frac{1}{c_0^2} \partial_t^3 \bM \cdot
\int_{\bx' \in \mathbb{D}_{\text{obj}}} 
\frac{[(\bx'\cdot \bnu)\bn - (\bx'\cdot \bn)\bnu]}{|\bx'- \bx_{\text{R}}|^2} \, \text{d}V
\right]
\end{split}
\end{align}
and 
\begin{align}
\label{eq:de_Born_td_QS}
\begin{split}
d_{\text{e}}^{\text{QS}}(t) &= 
-\frac{\mu_0A}{4\pi} 
\left[\frac{\sigma}{3} \partial_t \bM \cdot 
\int_{\bx' \in \mathbb{D}_{\text{obj}}} 
\frac{\bx'\times \bp_1}{|\bx'- \bx_{\text{R}}|^3}\,\mathrm{d}V \right. \\
&+\left.
\frac{\varepsilon_{\text{r}}-1}{3}\varepsilon_0\partial_t^2 \bM \cdot 
\int_{\bx' \in \mathbb{D}_{\text{obj}}} 
\frac{\bx'\times \bp_1}{|\bx'- \bx_{\text{R}}|^3}\,\mathrm{d}V \right. \\
&+ \left.
\varepsilon_0 \partial_t^2 \bM \cdot 
\int_{\bx' \in \mathbb{D}_{\text{obj}}} 
\frac{\bnu \times \bn}{|\bx'- \bx_{\text{R}}|^2}\,\mathrm{d}V \right],
\end{split}
\end{align}
for $t>0$ explicitly showing that time variations of the magnetization are received without any propagation delay in the quasi-static limit. 
We observe that for a magnetic field measurement, the conductivity and permittivity are present in the intermediate-field contribution to the signal ($1/\text{distance}^2$ term), while for an electric field measurement the dielectric properties of the ball show up in the near field contribution to the signal ($1/\text{distance}^3$ term). 

As we move away from the ball, the travel time $\tau$ will obviously increase. The above quasi-static signal models remain valid, however, provided that  $|s|\tau \ll 1$ for all $\bx' \in \mathbb{D}_{\text{obj}}$. Obviously, the quasi-static signal models can no longer be used as soon as this inequality is not satisfied.  

Finally, for later convenience we write the quasi-static signals as 
\begin{align}
\label{eq:d_h_qs_rew}
d_{\text{h}}^{\text{QS}}(t) &= 
 \partial_t \bM \cdot \ba_1^{\text{h}}(\bx_{\text{R}}) 
+  \frac{\sigma \mu_0}{3} \, \partial_t^2 \bM \cdot \ba_2^{\text{h}}(\bx_{\text{R}}) \nonumber \\
&+ \frac{\varepsilon_{\text{r}}-1}{3} \frac{1}{c_0^2}\, \partial_t^3 \bM \cdot \ba_2^{\text{h}}(\bx_{\text{R}})
\intertext{and}
\label{eq:d_e_qs_rew}
d_{\text{e}}^{\text{QS}}(t) &= 
\frac{\sigma}{3} \partial_t \bM \cdot \ba_1^{\text{e}}(\bx_{\text{R}}) 
+  \frac{\varepsilon_{\text{r}}-1}{3}\varepsilon_0 \, \partial_t^2 \bM \cdot \ba_1^{\text{e}}(\bx_{\text{R}}) \nonumber \\
&+  \varepsilon_0 \, \partial_t^2 \bM \cdot \ba_2^{\text{e}}(\bx_{\text{R}}),
\end{align}
 where the expressions for the expansion vectors $\ba_{k}^{\text{e,h}}(\bx_{\text{R}})$, $k=1,2$, are easily obtained from (\ref{eq:dh_Born_td_QS}) and (\ref{eq:de_Born_td_QS}).

%
% FIGURE CONFIGURATION
%
\begin{figure}[tbp]
\centering
\includegraphics[width=0.45\textwidth]{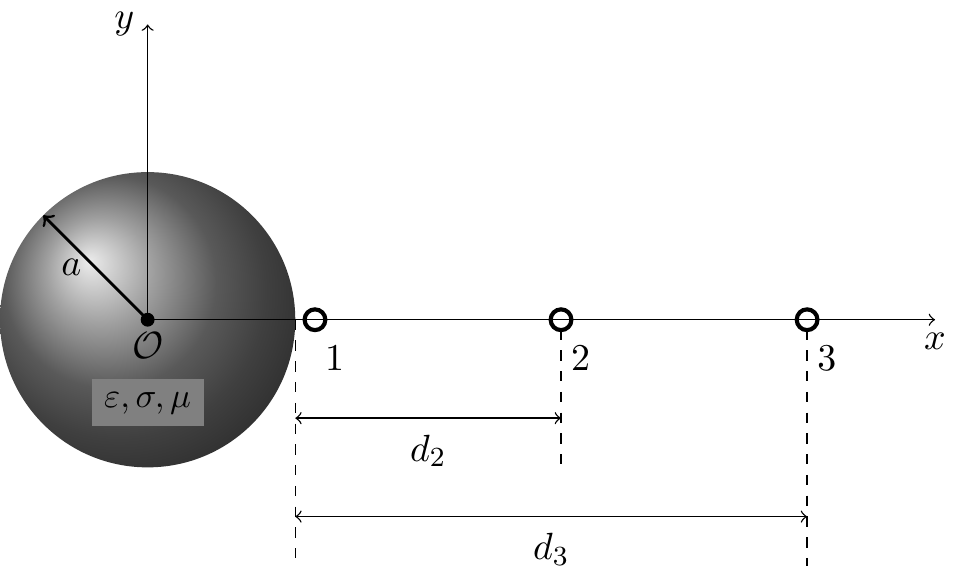}
\caption{Free induction decay signal measurement setup. A homogeneous ball with a radius $a$, centered at the origin has a permittivity $\varepsilon$, conductivity $\sigma$, and permeability $\mu$. Relaxation times $T_1$ and $T_2$.}
\label{fig:config}
\end{figure}

%
% TABLE MEDIUM PARAMETERS
%
\begin{table}[tbp]
\caption{Dielectric medium parameters of white matter for different background fields~\cite{website, Gabriel}}
\label{tab:EMpar}
\centering
\begin{tabular}{l c c c c c}
\toprule
$B_0$~[T]              & 1.5  & 3    & 7    & 11.2                   \\
\midrule
$\sigma~[\text{S/m}]$ & 0.3 & 0.3 & 0.4 & 0.5 \\
$\varepsilon_{\text{r}}$ & 68 & 53 & 44 & 41 \\
\bottomrule
\end{tabular}
\end{table}

\section{Simulations}
\noindent
To test the validity of our signal models and to study the influence of the permittivity and conductivity of the ball on these signals, we consider the configuration illustrated in Fig.~\ref{fig:config}. In this configuration, all geometrical parameters are fixed and wavelength independent, since we want to investigate this setup in MR scanners with different background fields. In particular, the radius of the ball is set to $a=2.5$~cm, and we use three receivers located on the $x$-axis to measure the various field responses. With Receiver~1 we carry out surface measurements, and in our simulations this receiver is located at a distance $d_1=2.5 \cdot 10^{-6}$~cm from the ball. Receiver~2 is located at a distance $d_2=25$~cm from the ball and, finally, Receiver~3 is located at a distance $d_3=50$~cm from the ball. All three receivers are loops that have a circular surface area with a radius of 2~cm. When we carry out a magnetic field measurement (emf), the loop is oriented in the $x$-direction ($\bnu=\bi_x$), while for an electric field measurement (mmf) we orient the loop in the $z$-direction ($\bnu=\bi_z$). The signal models will be evaluated for background fields of 1.5~T, 3~T, 7~T, and 11.2~T. The ball that we consider consists of white matter and its conductivity and relative permittivity values at the Larmor frequencies that correspond to these background fields are listed in Table~\ref{tab:EMpar}. In all cases, the relative permeability is taken to be equal to one. For the relaxation times of white matter we take those of a 3T background field, $T_1=900$~ms and $T_2=75$~ms \cite{bojorquez2016}, and we use these values for all background fields under consideration.

The signals that we receive are free induction decay (FID) signals as generated by the time-varying magnetization 
\begin{align}
\label{eq:FIDMx}
M_x(t) &= \phantom{-}M^{\text{eq}} e^{-t/T_2} \cos(\omega_0 t) \\
\label{eq:FIDMy}
M_y(t) &= -M^{\text{eq}} e^{-t/T_2} \sin(\omega_0 t) 
\intertext{and}
\label{eq:FIDMz}
M_z(t) &= M^{\text{eq}} (1-e^{-t/T_1}),
\end{align}
where $\omega_0=\gamma B_0$ is the Larmor frequency, $T_1$ and $T_2$ are the longitudinal and transverse relaxation times, respectively, and $M^{\text{eq}}$ is the equilibrium magnetization. For a proton spin density $\rho = 6.69\cdot 10^{28}~\text{m}^{-3}$ (water) and at $T=310.15$~K, the equilibrium magnetization evaluates to $M^{\text{eq}} \approx 0.0031 B_0$. The above components of the magnetization form the solution of the Bloch equation with initial condition $\bM(0)=M^{\text{eq}}\bi_x$. For $t>0$, the above solution describes how the magnetization relaxes back to its equilibrium $\bM=M^{\text{eq}}\bi_z$ as time increases.  

\subsection{Validating the Born approximation}
\begin{figure}[tbp]
\centering
\includegraphics[width=0.5\textwidth,trim={0.5cm 1.5cm 1cm 1.3cm}]{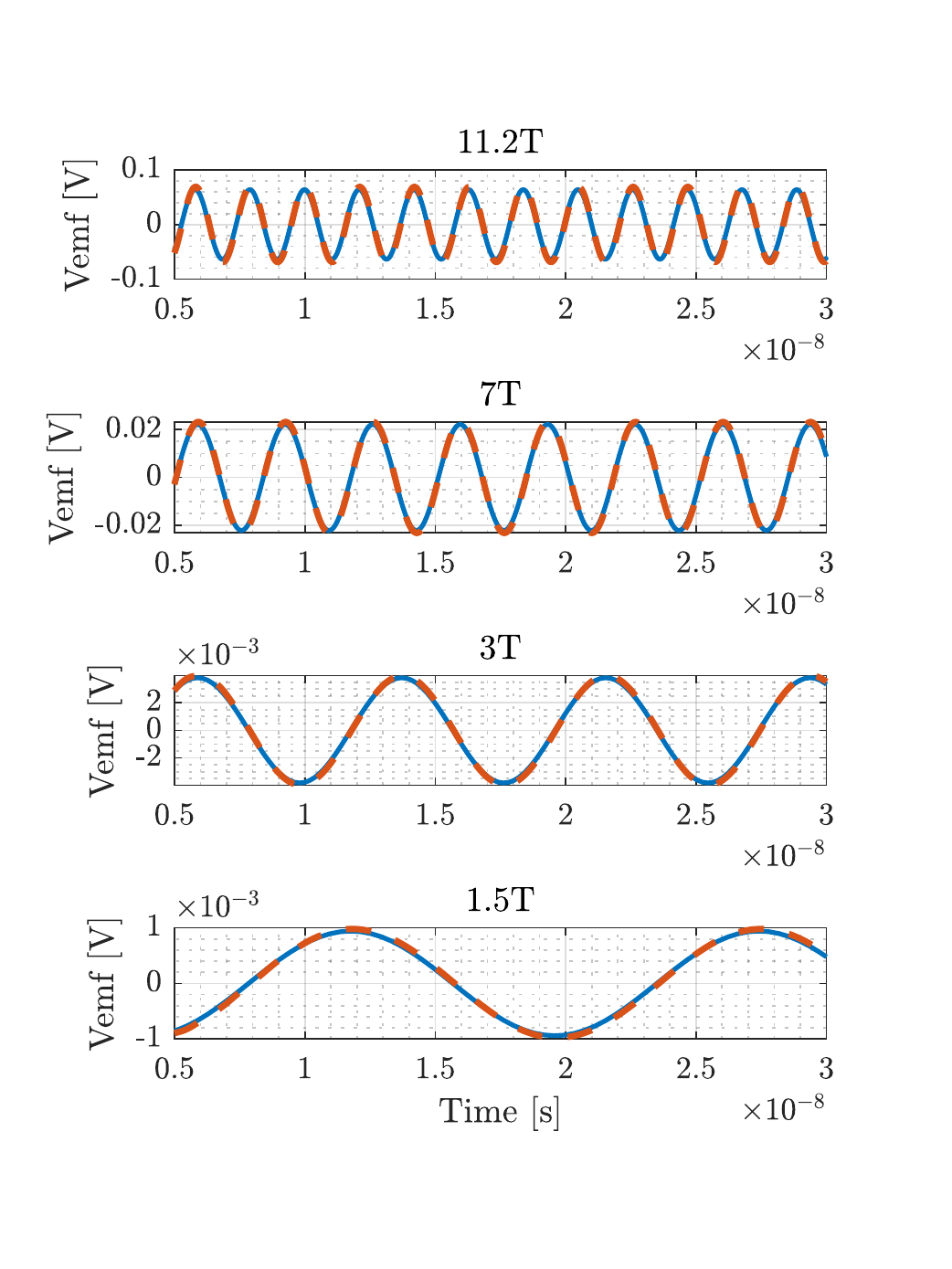}
\caption{Validation of the Born approximation for various background fields. Dashed (red) lines: $V_{\text{emf}}$ as determined from the magnetic field of the FDTD simulation. Solid (blue) line: the signal model of (\ref{eq:dh_Born_td}). The receiver is located at $\bx_\text{R}=[3.2,0.0,0]$~cm and the medium parameters of the ball are listed in Table~\ref{tab:EMpar}.}
\label{fig:Born}
\end{figure}

Before we carry out our signal analysis, we first validate the Born approximation for all background fields under consideration, since our signal models are based on this approximation. Specifically, we compute the time-domain electromagnetic field due to the magnetization given by (\ref{eq:FIDMx}) -- (\ref{eq:FIDMz}) using an in-house UPML-FDTD code. In our FDTD model, the conductivity and permittivity values of the ball at the various Larmor frequencies are selected according to Table~\ref{tab:EMpar}. Subsequently, we use the computed FDTD field responses to determine the electromotive force $V_{\text{emf}}$ as measured by a receiver located at $\bx_\text{R}=[3.2,0,0]$~cm. The dashed lines in Fig.~\ref{fig:Born} show the resulting signals for various background fields. The solid lines in this figure depict the signal model of (\ref{eq:dh_Born_td}) at the same receiver location and for the same background fields. This latter model is based on the quasi-static Born approximation  (\ref{eq:Born})  and (\ref{eq:Eb_QS}), while obviously no such approximation has been applied in our FDTD simulations. From Fig.~\ref{fig:Born} we observe that the signals based on FDTD modeling and the signals based on the quasi-static Born approximation overlap thereby validating that for this configuration and for all background fields of interest, the Born approximation indeed provides us with an accurate signal description.    

%
% FIGURE Born validation
%

\subsection{Quasi-Static Signal Analysis}
\begin{figure}[tbp]
\centering
\includegraphics[width=0.5\textwidth,trim={0.5cm 1.5cm 1cm 1.3cm}]{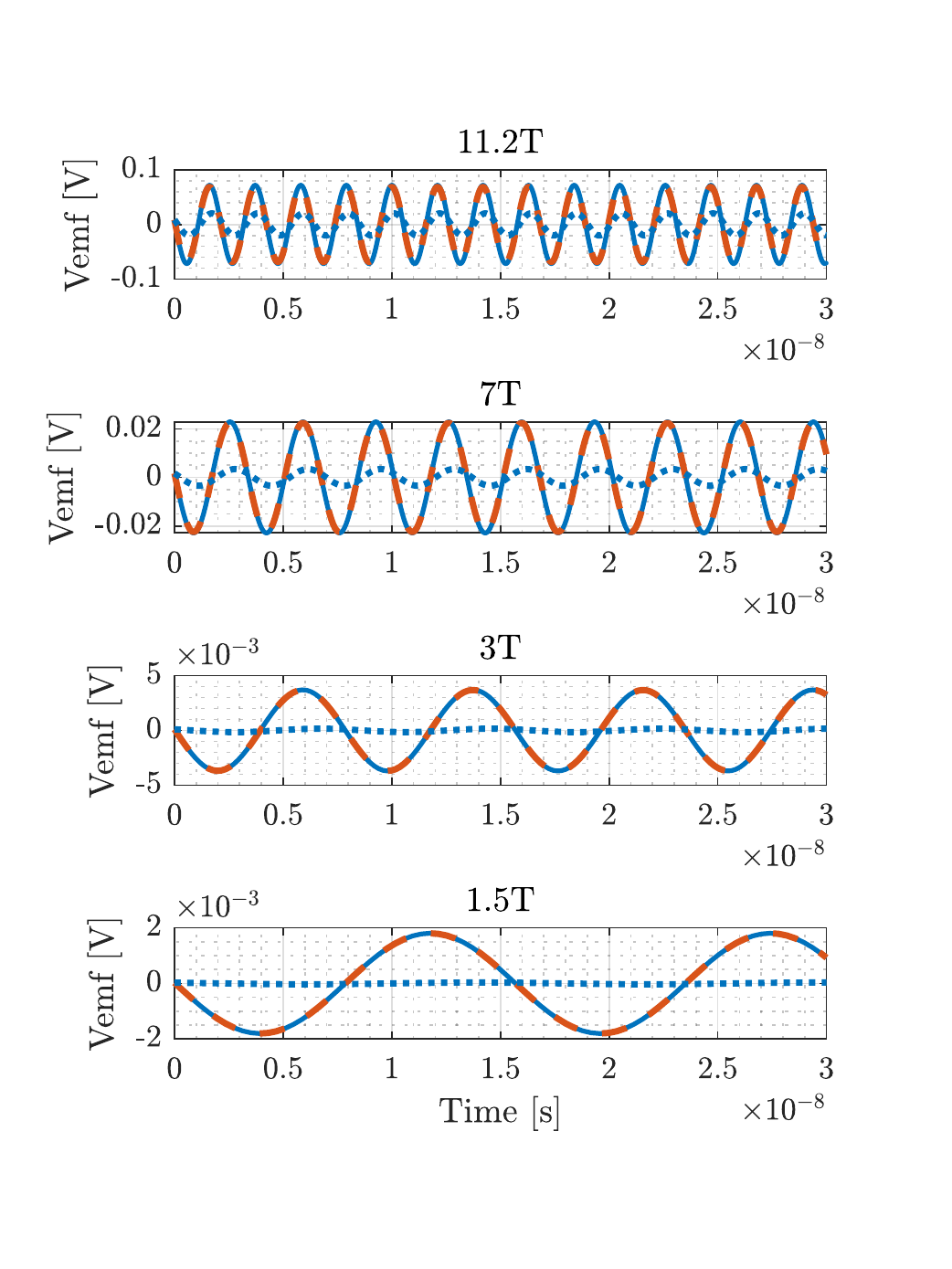}
\caption{Electromotive force at Receiver 1 for various background field strengths and a ball of white matter. The dielectric parameters are listed in Table~\ref{tab:EMpar}. Solid line: full wave signal model of (\ref{eq:dh_Born_td}); dashed line: quasi-static signal model of (\ref{eq:dh_Born_td_QS}); dotted line: sum of the last two terms on the right-hand side of (\ref{eq:dh_Born_td_QS}).}
\label{fig:rec1}
\end{figure}

\begin{figure}[tbp]
\centering
\includegraphics[width=0.5\textwidth,trim={0.5cm 1.5cm 1cm 1.3cm}]{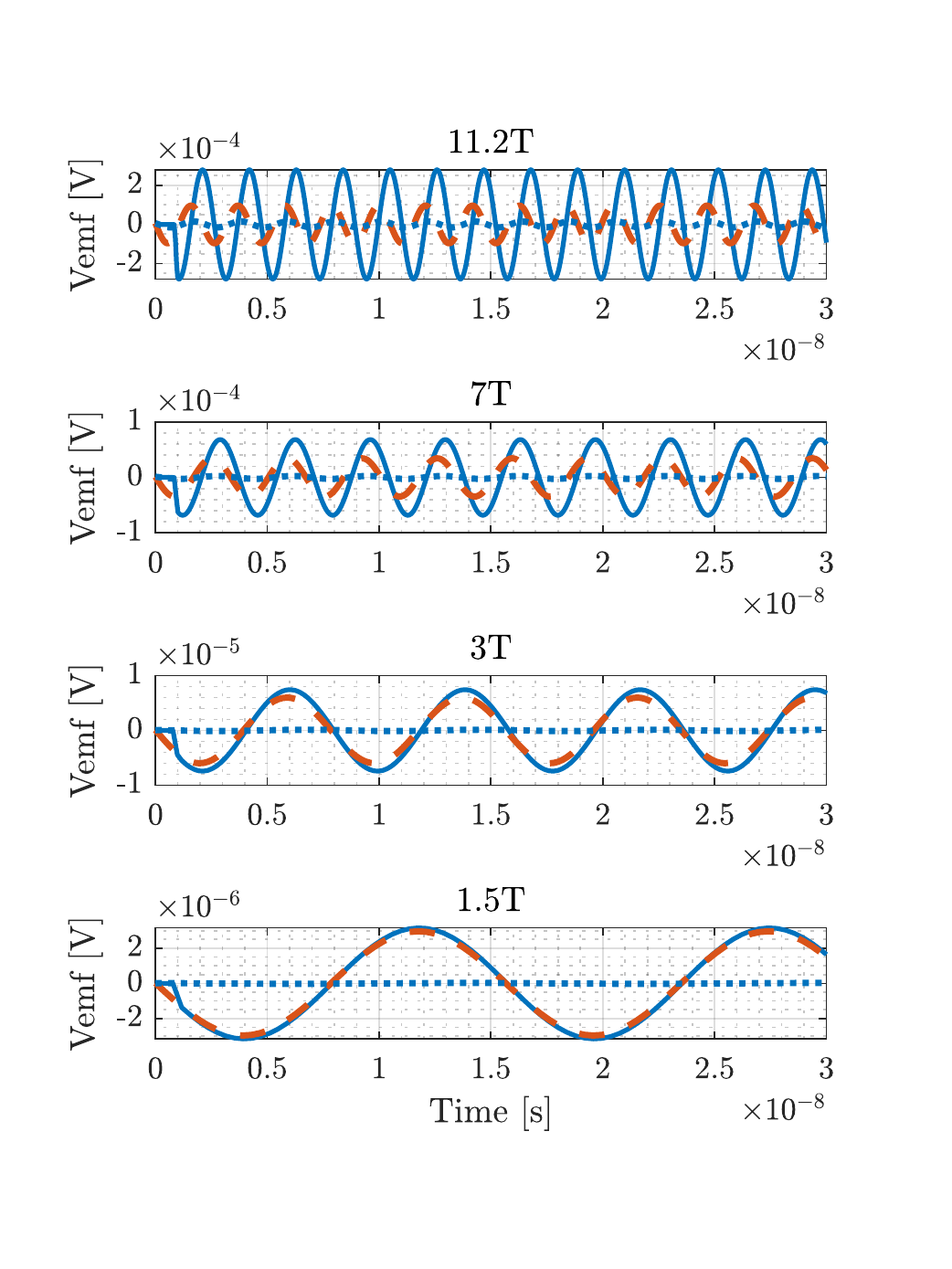}
\caption{Electromotive force at Receiver 2 for various background field strengths and a ball of white matter.  The dielectric parameters are listed in Table~\ref{tab:EMpar}.  Solid line: full wave signal model of (\ref{eq:dh_Born_td}); dashed line: quasi-static signal model of (\ref{eq:dh_Born_td_QS}); dotted line: sum of the last two terms on the right-hand side of (\ref{eq:dh_Born_td_QS}).}
\label{fig:rec2}
\end{figure}

\begin{figure}[htbp]
\centering
\includegraphics[width=0.5\textwidth,trim={0.5cm 1.5cm 1cm 1.3cm}]{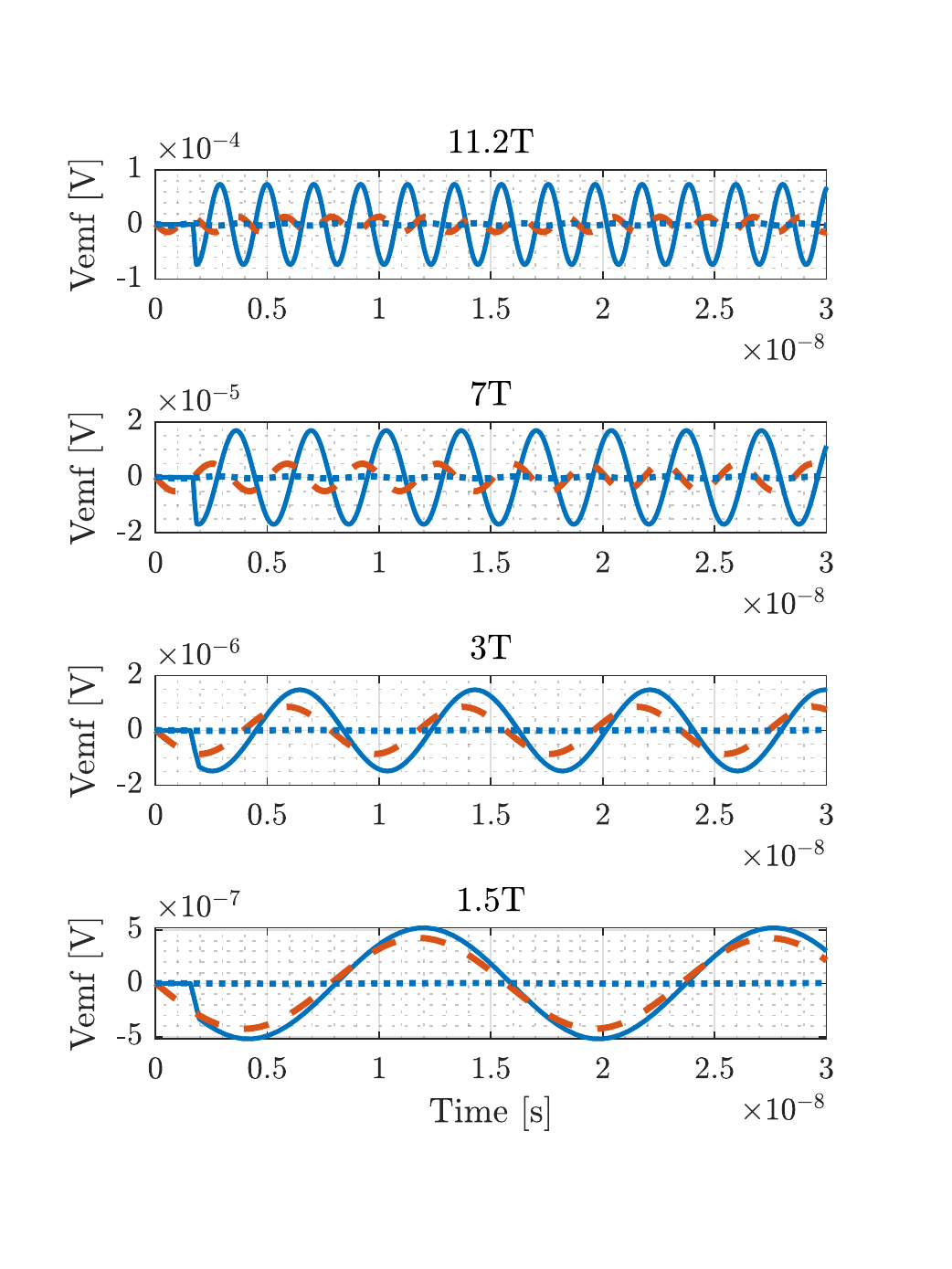}
\caption{Electromotive force at Receiver 3 for various background field strengths. The dielectric parameters are listed in Table~\ref{tab:EMpar}.  Solid line: full wave signal model of (\ref{eq:dh_Born_td}); dashed line: quasi-static signal model of (\ref{eq:dh_Born_td_QS}); dotted line: sum of the last two terms on the right-hand side of (\ref{eq:dh_Born_td_QS}).}
\label{fig:rec3}
\end{figure}

In the Laplace-domain, the quasi-static signal models hold provided that the condition $|s|\tau \ll 1$ is satisfied for all $\bx'\in \mathbb{D}_{\text{obj}}$ and all frequencies $s$ of interest. For the FID signals as generated by the magnetization of (\ref{eq:FIDMx}) -- (\ref{eq:FIDMz}), the Larmor frequency is the only non-vanishing oscillation frequency, and we can set $s=\text{j}\omega_0$ in the above condition to obtain the quasi-static requirement that $2\pi \lambda_0^{-1} |\bx'- \bx_{\text{R}}| \ll 1$ should hold for all $\bx'\in \mathbb{D}_{\text{obj}}$, where $\lambda_0$ is the wavelength in free space. Introducing the maximum distance $d_{\text{max}} = \underset{\bx'\in \mathbb{D}_{\text{obj}}}{\text{max}} |\bx'- \bx_{\text{R}}|$, the quasi-static condition is satisfied if $2\pi d_{\text{max}}/\lambda_0  \ll 1$.  Table~\ref{tab:par_dist} lists $2\pi  d_{\text{max}}/\lambda_0$ for the three receivers mentioned above and for different background fields. From this table, we expect the quasi-static approximation to hold for Receiver~1 and essentially all background fields under consideration. For Receiver~2, the quasi-static signal models are expected to hold for 1.5~T and possibly 3~T background fields, while for Receiver~3 the quasi-static field approximation possibly holds at 1.5~T only.
Figures~\ref{fig:rec1} -- \ref{fig:rec3} show the full wave signal model of (\ref{eq:dh_Born_td}) (solid line) and the quasi-static signal model of  (\ref{eq:dh_Born_td_QS}) (dashed line) for the electromotive force $V_{\text{emf}}$ at the three receivers of Fig.~\ref{fig:config}. Since a quasi-static electromotive or magnetomotive force signal analysis leads to the same conclusions, we present results for the electromotive force only. 

From Figs.~\ref{fig:rec1} -- \ref{fig:rec3} we observe that the quasi-static parameters of Table~\ref{tab:par_dist} quite accurately predict when a quasi-static signal model can be used. Specifically, for Receiver~1 the value of $2\pi d_\text{max}/\lambda_0 $ is at or below 0.5 for all background fields and Fig.~\ref{fig:rec1} shows that the full and quasi-static signals essentially overlap. For Receiver~2, however, we observe that the quasi-static model overlaps with the full-wave model for a background field of 1.5~T, but starts to deviate from the full-wave model for a background field of 3~T. For even higher background fields the quasi-static model is no longer valid, which is consistent with  Table~\ref{tab:par_dist}, since $2\pi d_\text{max}/\lambda_0 $ is larger than one in this case. These results indicate that the quasi-static signal model coincides with the full-wave model as long as $2\pi d_\text{max}/\lambda_0 \leq 0.5$. This observation is consistent with the full-wave and quasi-static signal models for Receiver~3 shown in Fig.~\ref{fig:rec3}. In this case, the quasi-static signal model already deviates from the full-wave model for a background field of 1.5~T for which we have $2\pi d_\text{max}/\lambda_0 \approx 0.74$. For higher background fields the quasi-static signal approximation definitely does not hold at Receiver~3 and we have to resort to the full-wave model of  (\ref{eq:dh_Born_td_QS})  in this case. 
  
Finally, the dotted lines in Figs.~\ref{fig:rec1} -- \ref{fig:rec3} show the contribution of the conductivity and permittivity terms (the last two terms on the right-hand side of (\ref{eq:dh_Born_td_QS})) to the total quasi-static signal (\ref{eq:dh_Born_td_QS}). We also observe that the contribution of these terms is small for lower background fields, but increases as the background field strength increases. Consequently, the conductivity and permittivity of the ball can be retrieved from a quasi-static electromotive force measurement, provided the SNR of the signals and the background field strengths are sufficiently large and the quasi-static field approximation holds. Another option is, of course, to use an electric field measurement (magnetomotive force measurement) as a basis for conductivity and permittivity retrieval, since for such a measurement these quantities contribute to the signal via the near-field as opposed to an electromotive force measurement, where the medium parameters contribute to the signal via the intermediate field.

%
% FIGURE Vemf at receiver 1
%

\begin{table}[tbp]
\caption{Background fields and normalized distances}
\label{tab:par_dist}
\centering
\begin{tabular}{l c c c c c}
\toprule
$B_0$~[T]              & 1.5  & 3    & 7    & 11.2                  \\
\midrule
$\lambda_0~[\text{m}]$ & 4.69 & 2.35 & 1.01 & 0.63 \\
$2a/\lambda_0$ & 0.01 & 0.02 & 0.05 & 0.08 \\
\midrule
Rec.~1: $2\pi d_\text{max}/\lambda_0 $ & 0.07 & 0.13 & 0.31 & 0.5 \\
Rec.~2: $2\pi  d_\text{max}/\lambda_0 $ & 0.40 & 0.80 & 1.87 & 3.00 \\
Rec.~3: $2\pi  d_\text{max}/\lambda_0$ & 0.74 & 1.47 & 3.44 & 5.50 \\
\bottomrule
\end{tabular}
\end{table}
%
% FIGURE Vemf at receiver 2
%

%
% FIGURE Vemf at receiver 3
%

\subsection{Conductivity and Permittivity Retrieval}
Since the quasi-static signal models under the Born approximation are all valid for measurements carried out with Receiver~1 (surface measurement) and all background fields of interest, we now use these models at this receiver location to retrieve the conductivity and permittivity of the ball (white matter). 

Let us start with the signal model for a magnetic field measurement given by (\ref{eq:d_h_qs_rew}). Introducing the functions $d_1^{\text{h}}(t) = \partial_t \bM \cdot \ba_1^{\text{h}}(\bx_{\text{R}})$,  $d_2^{\text{h}}(t) = \mu_0\partial_t^2 \bM \cdot \ba_2^{\text{h}}(\bx_{\text{R}})$, and  $d_3^{\text{h}}(t) = c_0^{-2} \partial_t^3 \bM \cdot \ba_2^{\text{h}}(\bx_{\text{R}})$, we have 
\begin{equation}
\label{eq:reconH_1}
d_{\text{h;QS}}^{\text{Born}}(t) = 
d_1^{\text{h}}(t) + \frac{\sigma}{3} d_2^{\text{h}}(t) + \frac{\varepsilon_{\text{r}}-1}{3} d_3^{\text{h}}(t),
\end{equation}
for $t>0$. Similarly, for the electric field signal model we have 
\begin{equation}
\label{eq:reconE_1}
d_{\text{e;QS}}^{\text{Born}}(t) = 
\frac{\sigma}{3} d_1^{\text{e}}(t) + \frac{\varepsilon_{\text{r}}-1}{3} d_2^{\text{e}}(t) +  d_3^{\text{e}}(t),
\end{equation}
for $t>0$ with $d_1^{\text{e}}(t) = \partial_t \bM \cdot \ba_1^{\text{e}}(\bx_{\text{R}})$,  $d_2^{\text{e}}(t) = \varepsilon_0\partial_t^2 \bM \cdot \ba_1^{\text{e}}(\bx_{\text{R}})$, and  $d_3^{\text{e}}(t) = \varepsilon_0 \partial_t^2 \bM \cdot \ba_2^{\text{e}}(\bx_{\text{R}})$.

Subsequently, we introduce the time instances $t_n = (n-1)\Delta t$ for $n=1,2,...,N$ with $(N-1)\Delta t = T_{\text{obs}}$, where $T_{\text{obs}}$ is the length of the observation interval, and consider the above signals at these time instances to obtain 
\begin{align}
\label{eq:reconH_2}
\bd^{\text{h}} &=  \bd_1^{\text{h}} +  \frac{\sigma}{3} \bd_2^{\text{h}} + \frac{\varepsilon_{\text{r}}-1}{3} \bd_3^{\text{h}} 
 \intertext{and}
 \label{eq:reconE_2}
 \bd^{\text{e}} &= \frac{\sigma}{3} \bd_1^{\text{e}} + \frac{\varepsilon_{\text{r}}-1}{3} \bd_2^{\text{e}} +  \bd_3^{\text{e}},
\end{align}
where $\bd^{\text{h}} = [d_{\text{h;QS}}^{\text{Born}}(t_1), d_{\text{h;QS}}^{\text{Born}}(t_2), ..., d_{\text{h;QS}}^{\text{Born}}(t_N)]^{T}$ is an $N$-by-1 column vector and all other vectors in the above equation are defined similarly. 

\begin{figure}[tbp]
\includegraphics[]{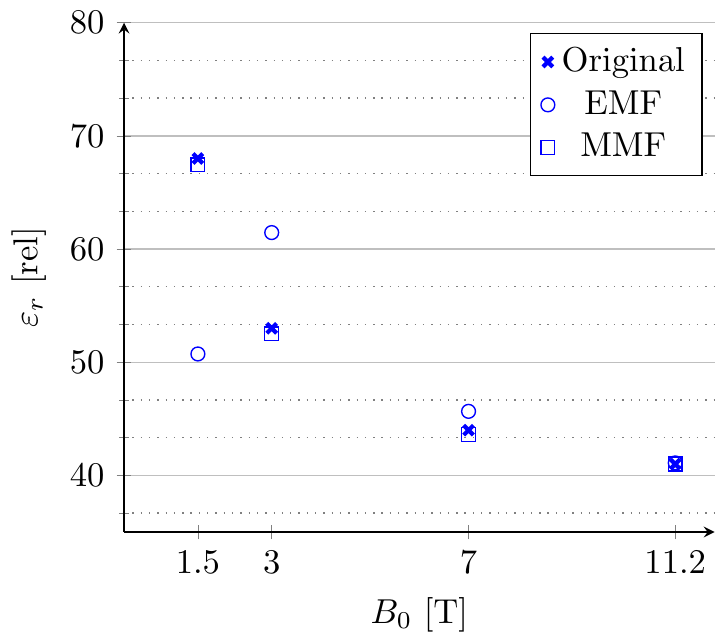}
\includegraphics[]{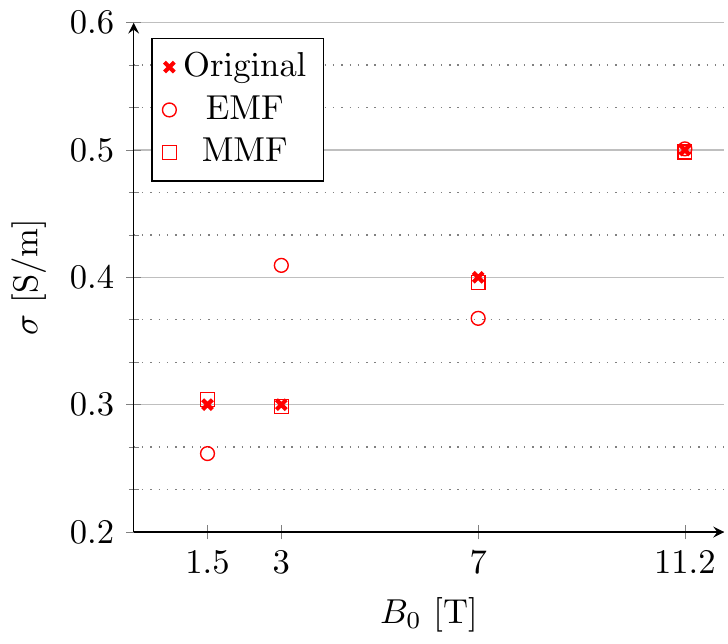}
\caption{Reconstructed permittivity (top) and conductivity (bottom) values using an EMF or MMF measurement for various field strengths.}
\label{fig:recons}
\end{figure}

Since we consider FID signals as generated by the magnetization of (\ref{eq:FIDMx}) -- (\ref{eq:FIDMz}), it immediately follows that the vector $\bd_1^{\text{h}}$ and $\bd_3^{\text{h}}$ and the vectors $\bd_2^{\text{e}}$ and $\bd_3^{\text{e}}$ are linearly dependent. Therefore, we consider the modified (scattered) data equations
\begin{equation}  
\label{eq:recon}
\tilde{\bd}^{\text{h}} = \bA^{\text{h}} \bc
\quad \text{and} \quad 
\tilde{\bd}^{\text{e}} = \bA^{\text{e}} \bc
\end{equation}
with $\bc=\frac{1}{3}[\sigma, \varepsilon_{\text{r}}-1]^{T}$, $\tilde{\bd}^{\text{h}} = \bd^{\text{h}} -  \bd_1^{\text{h}}$, $\tilde{\bd}^{\text{e}} =  \bd^{\text{e}} -  \bd_3^{\text{e}}$ and the matrices $\bA^{\text{h}}$ and $\bA^{\text{e}}$ have the column partitioning $\bA^{\text{h}}=(\bd_2^{\text{h}},\bd_{3}^{\text{h}})$ and $\bA^{\text{e}}=(\bd_1^{\text{e}},\bd_{2}^{\text{e}})$. Finally, noise is added to the data and we attempt to reconstruct the medium parameters as
\begin{equation}
\label{eq:min}
\bc^{\ast} = \underset{\bc}{\text{argmin}} 
\| \tilde{\bd}^{\text{h,e}}_{\text{n}} - \bA^{\text{h,e}} \bc \|_{2}^{2}
\end{equation}
where $\tilde{\bd}^{\text{h,e}}_\text{n} =\tilde{\bd}^{\text{h,e}} + \bn$ is the noisy data vector with $\bn$ the noise vector. With $T_0=2\pi/\omega_0$, we first take $T_{\text{obs}}=3T_0 =O(10^{-8,9})$~s in our minimization problem. Clearly, the exponential decay of the FID signal can be neglected in this case. With an SNR of 20~dB the conductivity and permittivity are determined by solving the corresponding least-squares problem (\ref{eq:min}) and the retrieved parameters are depicted in Fig.~\ref{fig:recons} along with the exact conductivity and permittivity values of white matter and for various background fields as listed in Table~\ref{tab:EMpar}. From this figure, we observe that for a magnetic field (emf) measurement, the error in the retrieved medium parameters decreases as the background field strength increases. At 1.5~T and 3~T, the medium parameters cannot be retrieved, but accurate medium parameters are obtained only at 11.2~T. Since the dielectric medium parameters contribute via the near field to a signal that is based on an electric field (mmf) measurement, we expect that these parameters can be reliably recovered for low and high background fields. From Fig.~\ref{fig:recons} we observe that this is indeed the case and similar to a magnetic field measurement, the reconstruction results improve as the strength of the background field increases. Finally, we mention that  we have repeated this experiment on an observation interval $T_{\text{obs}} = 3 T_2 = O(10^{-2})$~s and found similar results, showing that the electrical properties can also be recovered on an $O(10^{-2})$ time scale. 

\section{Discussion and Conclusions}
In this paper, we have presented full wave signal models for MRI field measurements. The models show that the magnetization and the induced electric scattering currents contribute to the measured signals, both weighted by their respective receive fields that are determined by the antenna that is used for reception. We have shown that to evaluate the models, the Green's tensors of the background medium must be known, along with the dielectric properties of the object and the magnetization within the excited part of the object must be known as well. For inhomogeneous background media, the Green's tensors can only be evaluated numerically in general, which may be a formidable task especially if electrically large objects are of interest. Moreover, for given dielectric medium profiles and a given magnetization, the electric field strength within the object must be computed, since it is required to determine the electric scattering source. In other words, apart from numerically computing the Green's tensors of the background medium, a forward problem for the electric field strength must be solved as well. Despite these computational bottlenecks, direct evaluation is possible in principle. Moreover, the models can be easily extended to include contrasts in permeability, but at the expensive of having to solve a coupled forward problem for the electric and magnetic field within the object of interest. 

To obtain explicit closed-form signal representations for electric and magnetic field measurements, we have considered a homogeneous ball that is embedded in free-space. Obviously, the Green's tensors of the background medium are now known and if the dielectric parameters and radius of the ball are ``sufficiently small," the quasi-static Born approximation applies meaning that the electric field within the ball may be approximated by the quasi-static background field, which is explicitly known. Obviously, there is now no need to solve a forward problem and the medium parameters show up explicitly in the resulting signal models. Travel time effects are still included in these models, since the quasi-static Born approximation applies to the electric field within the ball only. Quasi-static signal models may be obtained, however, for receiver locations for which travel time effects can be neglected. These signal models directly generalize the standard quasi-static models as normally used in MRI and clearly show how the dielectric parameters of the ball influence the measured signals. In fact, for FID signals obtained from an electric or magnetic field measurement, we demonstrated that the dependence of the signals on the medium parameters can even be used to retrieve these parameters. Specifically, we showed that for high background fields (7T and 11.2~T), electric (mmf) and magnetic (emf) field measurements allow for reliable parameter reconstructions, while at lower field strengths only electric field measurements can be used essentially because the dielectric parameters show up in the near-field of an electric field measurement and not in the near-field of a magnetic field measurement. 

Future work consists of experimentally validation of the full-wave and simplified quasi-static signal models, and its derived electrical properties reconstruction method.  
The simplified quasi-static models have their limitations, of course, and care should be taken when applying these models, since they are valid for a ball and under very special circumstances only (quasi-static field and Born approximation applies). However, the simplified models can be used to find the dielectric parameters of various tissue types and other materials using easily obtained FID signals or using other MRI signal acquisition schemes. 

Obviously, the full-wave models do not suffer from these limitations and allow us to determine how inhomogeneous dielectric tissue profiles influence the measured signals. Large-scale computations are required to determine the effects of the conductivity and permittivity profiles on the measured signals, but the models can potentially be used in a wide variety of applications ranging from for RF coil/antenna optimalization to reduce local SAR and optimize the SNR in specific imaging applications to the design of antenna arrays that maximize the sensitivity of the signals to the electrical properties as opposed to the magnetisation.

\appendix
\section{Expansion vectors for time-domain signal models}
The expansion vectors for a magnetic field measurement are given by
\allowdisplaybreaks
\begin{align}
\br^{\text{mg}}_0 &= \bp_1, \\
\br^{\text{mg}}_1 &= \bp_1 + \frac{1}{3} Z_0 \sigma \,
[(\bx' \cdot \bnu) \bn - (\bx'\cdot \bn)\bnu], \\
\br^{\text{mg}}_2 &= \bp_2 + \frac{1}{3} Z_0 \sigma \,
[(\bx' \cdot \bnu) \bn - (\bx'\cdot \bn)\bnu]\nonumber \\
&+ 
\frac{1}{3}(\varepsilon_{\text{r}}-1)
\frac{(\bx' \cdot \bnu) \bn - (\bx'\cdot \bn)\bnu}{|\bx'- \bx_{\text{R}}|}, \\
\br^{\text{mg}}_3 &= \frac{1}{3}(\varepsilon_{\text{r}}-1) 
\frac{(\bx' \cdot \bnu) \bn - (\bx'\cdot \bn)\bnu}{|\bx'- \bx_{\text{R}}|},
\end{align}
where $Z_0$ is the impedance of vacuum and $\varepsilon_{\text{r}}$ the relative permittivity of the ball. The expansion vectors for an electric field measurement are given by
with 
\begin{align}
\br^{\text{el}}_0 &= 
\frac{\sigma}{3} 
\bx'\times \bp_1, \\
\br^{\text{el}}_1 &= 
Y_0 \bq + \frac{\sigma}{3} 
\bx'\times \bp_1, \\
\br^{\text{el}}_2 &= Y_0 \bq + \frac{\sigma}{3} 
\bx'\times \bp_2, \\
\br^{\text{el}}_3 &= \frac{1}{3}Y_0 (\varepsilon_{\text{r}}-1) 
 \frac{\bx'\times \bp_2}{|\bx'- \bx_{\text{R}}|},
\end{align}
where $Y_0=(\varepsilon_0/\mu_0)^{1/2}$ is the admittance of vacuum, and 
\begin{equation}
\label{eq:def_q}
\bq = \bnu \times \bn + \frac{1}{3} (\varepsilon_{\text{r}}-1)  \frac{\bx'\times \bp_1}{|\bx'- \bx_{\text{R}}|}.
 \end{equation}
Note that these expansion vectors are independent of $s$, but do depend on the distance $|\bx' - \bx_{\text{R}}|$. 

%
% BIBLIOGRAPHY (order of appearance)
%

\end{document}